\documentclass[aps,prl,twocolumn,groupedaddress]{revtex4}
\usepackage[dvips]{graphics, color}
\usepackage{amsmath}
\usepackage{amssymb}
\usepackage{float}
\usepackage{bm,times}
\usepackage{graphicx}
\newcommand{\be}{\begin{equation}}
\newcommand{\ee}{\end{equation}}
\newcommand{\bea}{\begin{eqnarray}}
\newcommand{\eea}{\end{eqnarray}}

\begin{document}

\title{Cavity optomechanics with stoichiometric SiN films}

\author{D. J. Wilson}
\author{C. A. Regal}
\author{S. B. Papp}
\author{H. J. Kimble}
\address{Norman Bridge Laboratory of Physics 12-33, California Institute of Technology, Pasadena, California
91125}

\date{November 16, 2009}

\begin{abstract}

We study high-stress SiN films for reaching the quantum regime with mesoscopic oscillators connected to a room-temperature thermal bath, for which there are stringent requirements on the oscillators' quality factors and frequencies.  Our SiN films support mechanical modes with unprecedented products of mechanical quality factor $Q_m$ and frequency $\nu_m$ reaching $Q_{m} \nu_m \simeq2 \times 10^{13}$ Hz. The SiN membranes exhibit a low optical absorption characterized by Im$(n) \lesssim 10^{-5}$ at 935 nm, representing a 15 times reduction for SiN membranes.  We have developed an apparatus to simultaneously cool the motion of multiple mechanical modes based on a short, high-finesse Fabry-Perot cavity and present initial cooling results along with future possibilities.
\end{abstract}

\maketitle

Progress towards observing quantum fluctuations of a mesoscopic mechanical oscillator has accelerated with the recent successful use of cavity light forces to damp and cool mechanical motion \cite{Braginsky1970}.  This use of radiation pressure in optomechanical systems combined with cryogenic precooling will likely soon allow ground state cooling, with a variety of recent experiments demonstrating phonon occupations $\bar{n} < 100$ \cite{Marquardt2009note}.  Essential to this effort has been the development of ultra-high Q mechanical oscillators that are compatible with low-loss optical systems where radiation pressure dominates over photothermal effects.

An enabling advance in this field would be to push capabilities of optomechanics to an extreme where quantum limits could be achieved in the presence of a room temperature thermal bath.  Cryogen-free operation would greatly facilitate the integration of mesoscopic quantum mechanical oscillators into hybrid quantum systems.  For example, using cold atoms, mechanical oscillators could be coupled to atomic motional states or spin thus linking to a rich quantum optics toolbox \cite{Treutlein2007,Hammerer2009b}.  Via projective measurements utilizing atomic ensembles quantum effects could also be recognized without achieving full ground state cooling \cite{Hammerer2009a}.  Moreover, the capability for simultaneous coupling to multiple mechanical modes could enable the generation of multipartite entanglement among mechanical and optical modes \cite{Genes2008b}.

\begin{figure} \begin{center}
\includegraphics[width=\columnwidth]{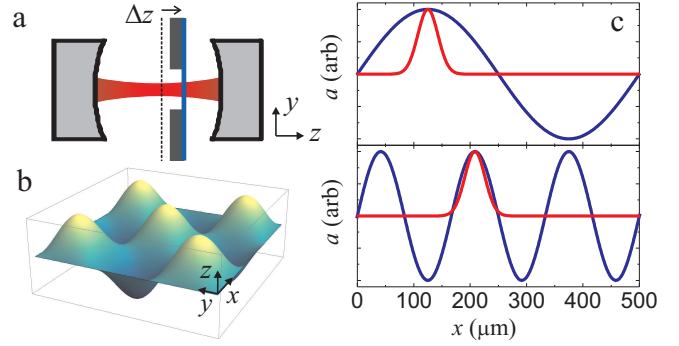} 
\caption{(a) Experiment schematic:  Membrane window in a short cavity.  (b)  Illustration of $(j,k)=(3,3)$ membrane mode
amplitude $a_{jk}$. (c) Optical mode $\psi(x,y)$ (red solid line) compared to membrane modes $(j,k)=(2,2)$ (top) and $(6,6)$ (bottom) for a 0.5 mm $\times$ 0.5 mm square membrane.} \label{schematic}
\end{center}
\end{figure}

A promising optomechanical platform introduced at Yale is a geometry in which a flexible SiN membrane with exceptional mechanical properties is coupled to a standard high-finesse Fabry-Perot cavity \cite{Zwickl2007,Thompson2007,Jayich2008}.  In this Letter we demonstrate that SiN membranes can be optimized to realize one of the key minimum requirements for approaching ground state cooling from room temperature, namely a mechanical quality factor $Q_m$ larger than the number of room temperature thermal phonons, i.e. $Q_m > \bar{n}_T=k_b T_{{\rm room}} / \hbar \omega_m$ \cite{Marquardt2007a}.  We realize an optomechanical system in which cavity cooling can be applied to multiple higher-order modes of these films.  This system is based on a short high finesse fabry-perot with a small mode waist (Fig. \ref{schematic}(a,c)) and displays low optical absorption and scattering at a wavelength of interest for cold-atom systems \cite{Treutlein2007,Hammerer2009b,Hammerer2009a}.  We analyze the prospects for accessing quantum effects from room temperature in this new regime.

SiN under tensile stress has been recognized for some time for its unusually low mechanical dissipation, particularly among amorphous materials \cite{Verbridge2006,Verbridge2008,Southworth2009,Thompson2007}. Recent optomechanical experiments with SiN membranes used the fundamental mode of a 1 mm low tensile stress film with $Q_m=1.1\times 10^6$ at $\omega_m=2 \pi\times 130$ kHz \cite{Thompson2007}.  In our experiments we use a SiN film in its stoichiometric form, Si$_3$N$_4$, which naturally has a large tensile stress of $T\sim1$ GPa.  We use sub-mm membranes and focus on the high-order modes (as illustrated in Fig. \ref{schematic}(b));  this along with the increased tension allows us to increase the resonant frequency of our mechanical mode significantly over the mode cooled in Ref. \cite{Thompson2007}.  This method of using tension to increase the frequency of mechanical modes while maintaining low dissipation has been recognized as an important tool for seeing quantum effects in a variety of mechanical systems \cite{Thompson2007,Corbitt2007,Verbridge2006,Regal2008}.

\begin{figure} \begin{center}
\includegraphics[width=\columnwidth]{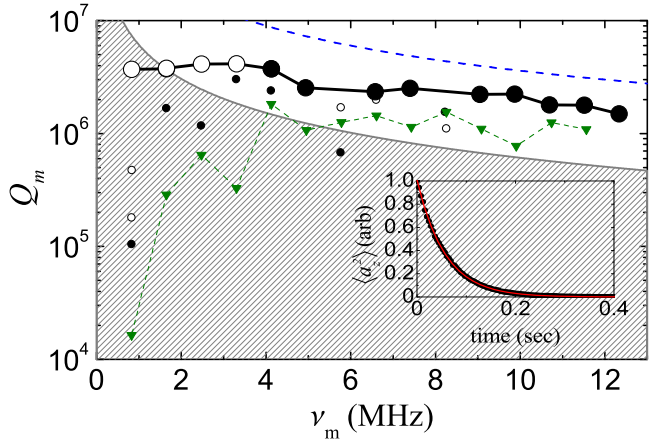} 
\caption{Mechanical quality factors of Si$_3$N$_4$ films measured as a function of resonant frequency for diagonal ($j=k$) modes.  Three different data sets are represented:  Membrane frame corners resting on a curved surface (open circles), membrane frame glued at three corners (closed circles), membrane frame glued over the full frame (green triangles).  The solid black line is a guide to the eye of the maximum $Q_m$ realized.  The grey shaded area represents a $Q$-frequency product less than $k_b T_{{\rm room}}/h = 0.6 \times 10^{13}$ Hz.  The dashed blue line is an estimate of the thermoelastic limit based upon an extension of Zener theory for a thin plate at room temperature \cite{theorynote,Verbridge2006,Zwickl2007}.  (inset) Representative mechanical ringdown for the (8,8) mode at 6.6 MHz.} \label{Qm}
\end{center}
\end{figure}

Specifically, we use 50 nm thick LPCVD nitride membranes from Norcada Inc in a square geometry of size $d\times d=0.5\times0.5$ mm.  We characterize the mechanical quality factor by monitoring the ringdown of the mechanical excitation as a function of time (Fig. \ref{Qm} (inset)).  To probe the excitation we use an etalon formed between the membrane and a partially reflective mirror.   The entire setup is placed under vacuum at $10^{-7}$ Torr.

In Fig. \ref{Qm} we plot the measured $Q_m$ as a function of mode frequency; these mode frequencies are consistent with $T=0.9$ GPa, a film mass density of $\rho = 2.7$ g/cm$^3$ \cite{Verbridge2006}, and the expectation of $\nu_m(j,k) = \omega_m(j,k)/2\pi=\sqrt{T/4\rho d^2}\sqrt{j^2+k^2}$, where $j$ and $k$ are mode indices.  The highest quality factor observed reaches $Q_m=4\times10^6$; the $Q$-frequency product ($Q_{m} \nu_m$) reaches $2\times10^{13}$ Hz and exceeds $1\times10^{13}$ Hz over a wide range of frequencies from 2 to 12 MHz.  The three sets of points in Fig. \ref{Qm} represent separate trials in which a new membrane was mounted in the apparatus in specific ways.  Variation between trials is likely due to mechanical coupling between the membrane and the mounting structure.  In addition to these three representative sets we have mounted and characterized dozens of other membranes.  Based upon our body of results, we infer that mounting techniques that utilize the least contact to the frame allow us to realize the highest quality factors.

Our $Q$-frequency products clearly extend above the grey-shaded region in Fig. \ref{Qm}, which represents a $Q$-frequency product less than $k_b T_{room}/h$.  High-stress Si$_3$N$_4$ was previously used to create membranes with $Q$-frequency products reaching $0.08 \times 10^{13}$ Hz and long nanostrings reaching $0.7 \times 10^{13}$ Hz \cite{Verbridge2006,Verbridge2008}.  Our measured products surpass these results and achieve a greater than 10 fold improvement over room-temperature optomechanical systems in which cavity cooling has been implemented thus far (see for example Refs. \cite{Thompson2007,Schliesser2008}).  Furthermore, our results illustrate that these amorphous oscillators approach the thermoelastic limit \cite{theorynote} and support $Q$-frequency products that rival the best observed at room temperature in single-crystal materials and diamond (see Ref. \cite{Lee2009} for a summary).

To cool the higher-order modes of these membranes, which have the best $Q$-frequency products, we must be able to selectively probe individual antinodes of these modes within a high-finesse cavity (Fig. \ref{schematic}(a,b)).  We create a small mode spot size by using a short Fabry-Perot cavity ($L=0.74$ mm) with small radius of curvature mirrors ($R_c=5$ cm). This results in a TEM$_{00}$ optical mode with a $1/e^2$ diameter of $2w_0=$71 $\mu$m at our chosen wavelength of $\lambda_0=935$ nm, which is a ``magic" wavelength for trapping of atomic cesium\cite{McKeever2003}.  In Fig. \ref{schematic}(c) this mode size is compared to the mechanical excitation, given by $a_{jk} (x,y)=a_z \sin{(j \pi x/d)} \sin{(k \pi y/d)}$.

\begin{figure} \begin{center}
\includegraphics[width=\columnwidth]{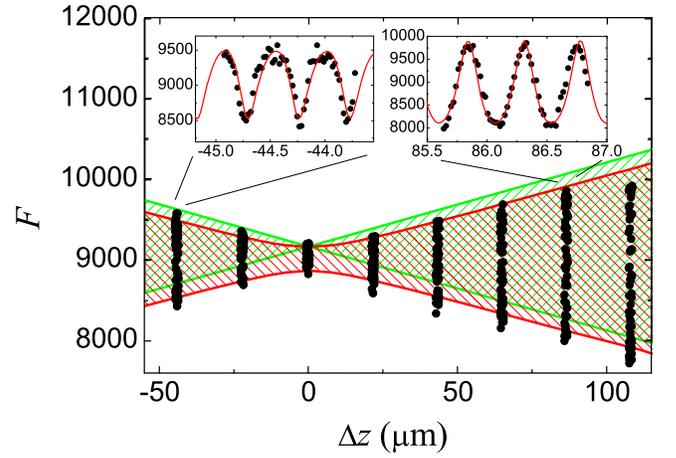} 
\caption{Measured cavity finesse $F$ as a function of membrane position $\Delta z$ about the center of the cavity (black circles).  Absent the membrane the finesse is $F_0 = 9160$.  The shaded regions represent the envelope of the calculated finesse variation for Im$(n)=0.6 \times 10^{-5}$ (red) and Im$(n)=0$ (green) for comparison. (Insets) Examples of the local dependence as a function of $\Delta z$ compared to theoretical curves for Im$(n)=0.6\times10^{-5}$.} \label{absorption}
\end{center}
\end{figure}

Another important feature of our optomechanical system is the absorption properties of our Si$_3$N$_4$ films.  We measure the absorption by studying the effect of the membrane on the cavity finesse \cite{Jayich2008}.  Fig. \ref{absorption} shows the finesse $F$ measured in transmission through the cavity as a function of membrane position $\Delta z$.  The measured $F$ is modulated as a function of membrane displacement, $\Delta z$, with a short periodicity of $\lambda_{0}/2$ set by the cavity standing wave (insets to Fig. \ref{absorption}), as well as on the larger scale of $L/2$ with respect to the cavity center ($\Delta z = 0$) with the envelope of modulation of $F$ exceeding the bare cavity finesse $F_{0}=9160$.  These effects can be understood in terms of the eigenmodes of the composite cavity with Im$(n)=0$, for which the ratio of intracavity field amplitudes $A_{l,r}$ on the left and right of the membrane varies periodically with $\Delta z$. For a membrane with reflectivity $R_{m}=1-\epsilon$ with $\epsilon \ll1$, the limiting values for the envelope of the finesse variation will be approximately $F_{l,r} \simeq F_{0} (1\pm 2\Delta z/L)$, while for $R_{m} \simeq 0$, the variation of finesse approaches zero. Our membrane with $R_{m}\simeq 0.18$ lies between these extremes, for which the numerical solution shown in Fig. \ref{absorption} has been obtained \cite{Wilson2009b}.

We have calculated the expected $F$ measured in transmission for arbitrary $\Delta z$ assuming a real index of $n=2.0$.
A one-parameter fit of the data to our model (Fig. \ref{absorption}) reveals absorption corresponding to an imaginary part of the index of Im$(n)$ $=0.6 \times10^{-5}$ for our films, assuming scattering losses are negligible (Fig. \ref{absorption}). The data with $|\Delta z| \gtrsim 50$ $\mu$m are more consistent with Im$(n)$ as high as $0.9\times10^{-5}$, which may be a manifestation of alignment effects.  These values are $15-25$ times lower than recently observed with low-stress nitride \cite{Zwickl2007,Jayich2008}, but consistent with studies that show that SiN can have absorption as low as Im$(n) < 10^{-6}$ in the near-IR \cite{Barclay2006,Inukai1994}.  Decreasing absorption is important for reaching the full quantum limit of motion detection \cite{Caves1981a} and is required for achieving the large linear coupling and high finesse required for strong coupling between one atom and a membrane \cite{Hammerer2009b}.

\begin{figure} \begin{center}
\includegraphics[width=\columnwidth]{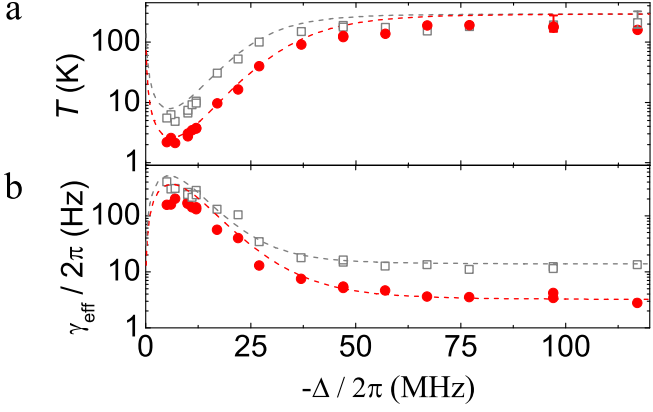} 
\caption{Demonstration of cavity cooling of the $(6,6)$ (red circles) and $(1,1)$ (grey squares) modes with frequencies and initial quality factors of $\nu_m=4.82$ MHz; $Q_m=1.5\times 10^6$ and $\nu_m=0.80$ MHz; $Q_m=5.8\times 10^4$, respectively, where $Q_m$ is measured via ringdown.  For these data the cavity output power at resonance is fixed at $P_{{\rm out}}^0=10$ $\mu$W.  The representative error bars show the uncertainty due to our knowledge of $g$. (a) Measured mode temperature (points) and theoretically expected cooling (lines) as a function of detuning $\Delta$ from the cavity resonance. (b) Corresponding measured effective mechanical linewidth $\gamma_{\rm eff}$ and theoretical expectation. } \label{cooling}
\end{center}
\end{figure}

In our high-finesse cavity we have realized radiation pressure cooling of a number of high-order membrane modes.  Figure \ref{cooling} demonstrates simultaneous cooling and damping of the $(1,1)$ and $(6,6)$ modes as a function of the detuning $\Delta=\omega_L-\omega_c$ of an input laser field, where $\omega_L$ and $\omega_c$ are the laser and cavity resonance frequencies respectively.  To cool the membrane we use light from a diode laser at 935 nm.  We probe the membrane's displacement by monitoring the amplitude modulation of the cooling light transmitted through the cavity off resonance.  We calculate the membrane motion for a mode at $\omega_m$ via $\langle a_z^2 \rangle_{\omega_m}=\frac{\langle i^2 \rangle_{\omega_m}}{i^2(\Delta)} \frac{(\gamma/2)^2}{g^2} \frac{1}{H(\omega_m,\Delta)}$ where $i(\Delta)$ is the dc photocurrent at detuning $\Delta$ and $\sqrt{\langle i^2 \rangle_{\omega_m}}$ is the rms photocurrent fluctuation obtained by integrating the noise spectrum over a lorentzian peak centered at $\omega_m$.  $H(\omega_m,\Delta)=\Delta^2 S_+ S_-$ is a dimensionless factor for the cavity response at arbitrary detuning, where $S_{\pm}=\gamma/[(\omega_m \pm \Delta)^2+(\gamma/2)^2]$. The membrane coupling is described by $g$, which for operation at optimal linear coupling is $0.85 \eta_{jk} \omega_c/L$.  The factor of 0.85 is related to the reflectivity of the membrane \cite{Zwickl2007}, whereas  $\eta_{jk}=|\int{\int{dx \, dy \, \psi(x,y) \, a_{jk}(x,y)/a_z}}|$ accounts for the mode overlap (Fig. \ref{schematic}(c)) \cite{Gillespie}.  Here $\psi(x,y)=(2/\pi w_0^2){\rm exp}(-2r^2/w_0^2)$, and $a_{jk}(x,y)$ is the membrane mode function discussed earlier.  We operate near the center of a (6,6) antinode at a distance of $(\delta x,\delta y) \simeq (45 \hspace{4pt} \mu {\rm m}, 120 \hspace{4pt} \mu {\rm m})$ from the center of the (1,1) mode; here $\eta_{66} \simeq 0.63$ and $\eta_{11} \simeq 0.69$, where these values are known at the $10\%$ level.  The mode temperature in Fig. \ref{cooling}(a) is then extracted from the integrated mechanical response via $\langle a_z^2 \rangle_{\omega_m} = k_b T /\ m_{e} \omega_m^2$, where the effective mass $m_e$ is $\frac{1}{4}$ the physical mass for all modes.  The temperature measured at large detuning agrees with our expectation for room temperature operation to within the uncertainty.

We can compare the observed cooling to our expectation based upon a simple cavity-cooling model relevant in the limit of weak coupling (lines in Fig. \ref{cooling}).  We operate in a regime where the optical linewidth $\gamma=2 \pi \times 25$ MHz (FWHM for $F=8100$) is somewhat larger than $\omega_m$, in between the so-called bad and good cavity limits \cite{finessenote,Marquardt2007a,WilsonRae2007}.  Here the full expression for the cooling rate is given by  $\gamma_{{\rm opt}} = g^2 x_{zp}^2 |\alpha|^2 (S_+-S_-)$ where $x_{zp}=\sqrt{\hbar/2 m_e \omega_m}$ and $|\alpha|^2$ is the effective intracavity photon number, which we estimate as $\frac{2 P_{{\rm out}}^0}{\hbar \omega_0} \frac{\gamma/2}{(\gamma/2)^2+\Delta^2}$.  The calculated effective damping rate $\gamma_{{\rm eff}}=\gamma_m+\gamma_{{\rm opt}}$ and the expected temperature in this regime $T=T_{{\rm room}} \gamma_m/\gamma_{{\rm eff}}$ follow the trend of the data.

 We have investigated the cooling theoretically achievable for our realized parameters.  Strictly, full ground state cooling requires $\bar{n}_T \gamma_m \ll \gamma \ll \omega_m $, i.e. fully resolved sidebands and an oscillator with a large enough $Q_m$ to allow the required damping.  For our results it is most relevant to consider instead the case $ \bar{n}_T \gamma_m \lesssim \gamma \sim \omega_m$ and track the results into a regime of strong cooling, characterized by $g_{{\rm eff}}=g x_{zp} \alpha > \gamma$ \cite{Marquardt2007a,Dobrindt,Groblacher2009}.  We thus calculate the expected phonon occupation $\bar{n}$ based on an exact solution to the coupled equations of motion \cite{Marquardt2007a,Genesnote} and integrate over the spectral function $S_{bb}(\omega)=\int dt e^{i\omega t} \langle \hat{b}^\dagger(t) \hat{b} \rangle$ where $\hat{b}^{\dagger}+\hat{b}=\hat{z}$.

\begin{figure} \begin{center}
\includegraphics[width=\columnwidth]{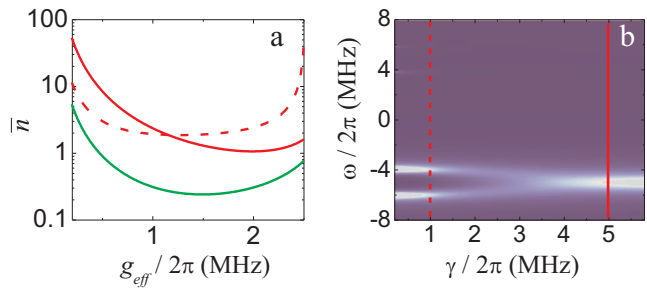}
\caption{(a) Calculated achievable phonon number as a function of coupling strength $g_{{\rm eff}}$ for parameters described in the text.  The solid line is for a cavity with $\gamma=2 \pi \times 5$ MHz and the dashed line for $\gamma=2 \pi \times 1$ MHz.  The green line would result  for $\gamma=2 \pi \times 5$ MHz upon increasing $Q_m$ to $4\times10^7$, which may be possible for a thinner SiN film.  (b)  Emergence of normal-mode splitting as the cavity linewidth is varied for fixed mechanical frequency $\omega_m=2\pi \times 5$ MHz and coupling $g_{{\rm eff}}=2\pi \times 1$ MHz.  The exact spectrum $S_{bb}(\omega)$ is displayed in this density plot where zero weight corresponds to the blue shading and white to maximum mechanical response.  The lines mark the position of the parameters used in (a).} \label{theory}
\end{center}
\end{figure}

Figure \ref{theory}(a) shows the calculated phonon number as a function of the cooling strength for the realistic parameters:  $Q_m=4 \times 10^6$, $\nu_m=5$ MHz, $\Delta=-\omega_m$, $\bar{n}_T=k_b T_{{\rm room}}/\hbar \omega_m=1.2 \times 10^6$, and $\gamma=2 \pi \times 5$ and 1 MHz (see caption).  Figure \ref{theory}(a) shows reaching $\bar{n}\sim 1$ from room temperature is theoretically achievable for demonstrated parameters before reaching the static bistability point ($g_{{\rm eff}}=\omega_m/2$ for $\gamma \ll \omega_m$) \cite{Dorsel1983}.  Another outstanding goal is achieving strong coupling between a ground state mechanical resonator and the cavity field \cite{Marquardt2007a,Dobrindt,Groblacher2009}.  Figure \ref{theory}(b) illustrates the spectral function $S_{bb}(\omega)$ as a function of the cavity linewidth for $g_{{\rm eff}} = 2 \pi \times 1$ MHz.  The normal-mode splitting indicative of strong coupling appears for phonon occupancies near unity.

Observing quantum effects of a mesoscopic oscillator coupled to an ambient thermal bath would be a significant advance.  Future work should address experimental challenges to achieving the occupations shown in Fig. \ref{theory}, such as reducing laser phase noise and mitigating the thermal noise of the Fabry-Perot cavity substrates \cite{Kimble2002}.  However, to implement a full range of quantum protocols with these oscillators one must achieve occupations $\bar{n} \ll 1$.  This will require higher $Q$-frequency products or lower initial thermal occupation, as illustrated by the green line in Fig. \ref{theory}(a).   Realizing even higher room temperature $Q$-frequency products than those shown in Fig. \ref{Qm} will be a subject of future investigation, where we note one limitation to consider is thermoelastic dissipation (dashed line in Fig. \ref{Qm}).  Further, we note that recent results in Ref. \cite{Southworth2009} indicate that, unlike amorphous SiO$_2$, the dissipation in SiN films decreases monotonically from room temperature down to $\sim$100 K, where liquid nitrogen can be used for cooling.

\acknowledgements{We thank Oskar Painter for valuable insights, especially into the absorption properties of Si$_3$N$_4$, and D. E. Chang for helpful discussions. This work was supported by the NSF and by IARPA via the ARO. CR acknowledges support from a Millikan Postdoctoral Fellowship; SP acknowledges support as fellow of the Center for the Physics of Information.}



\begin{thebibliography}{10}
\bibitem{theorynote}
Specifically our estimate makes use of Eqns. 4.16 - 4.19 in Ref. \cite{Norris}
  for $\omega \tau_0 \ll 1$. We estimate a thermal conductivity of 20 W/m/K
  from bulk nitride values, and we assume a linear expansion coefficient of
  $1.6 \times 10^{-6}$ K$^{-1}$, a Young's modulus of 270 GPa, and a Poisson
  ratio of 0.27 \cite{GadelhakCRC}. We note that uncertanties in these
  parameters can impact the calculation signficantly.
\end{thebibliography}

\end{document}